\documentclass[aps,prb,twocolumn,showpacs,preprintnumbers,amsmath,amssymb,superscriptaddress,noshowkeys]{revtex4-1}
\usepackage{txfonts}
\usepackage[dvips]{graphicx}
\usepackage{bm}
\usepackage{color}
\usepackage{ulem} 
\def\rnum#1{\expandafter{\romannumeral #1}} 
\def\Rnum#1{\uppercase\expandafter{\romannumeral #1}}

\newfont{\bg}{cmr10 scaled\magstep4}

\newcommand{\bigzerou}{\smash{\lower1.8ex\hbox{\bg 0}}}
\begin{document}
\title{Tuning of Weyl point emergence in multi-terminal Josephson junctions using quantum point contacts}
\author{Kento Takemura}
\email[E-mail me at: ]{takemura.k@opt.mp.es.osaka-u.ac.jp}
\affiliation{Department of Material Engineering Science,
Graduate School of Engineering Science, Osaka University,
1-3 Machikaneyama, Toyonaka, Osaka 560-8531, Japan}
\author{Mikio Eto}
\affiliation{Faculty of Science and Technology, Keio University, 3-14-1 Hiyoshi, Kohoku-ku, Yokohama 223-8522, Japan}
\author{Tomohiro Yokoyama}
\email[E-mail me at: ]{tomohiro.yokoyama@mp.es.osaka-u.ac.jp}

\date{\today}

\begin{abstract}
Multi-terminal Josephson junction with three or more superconductors is an attractive quantum system to emerge and tune exotic electronic states.
In four terminal Josephson junctions, the Weyl physics, namely topologically protected zero energy state, emerges without assuming any exotic materials.
In this study, we consider the four-terminal Josephson junction with the quantum point contact structures between the mesoscopic normal region and four superconducting terminals.
The quantum point contacts can tune electrically the number of conduction channels.
We theoretically investigate an effect of the increase of channels on the emergence of Weyl points.
The increase of channels causes the increase of Andreev bound states in the system, which increase the emergence probability of Weyl points.
When all terminals have two channels, the emergence probability is up to 17\%, which is about four times larger than that for all single channel junctions.
We consider the balance of the number of conduction channels in the four terminals.
When the number of channels is unbalanced, the increase of emergence probability is suppressed.
\end{abstract}
\maketitle

\section{INTRODUCTION}
\label{sec:intro}

In a conventional Josephson junction, one normal conductor is sandwiched by two superconductors.
In the normal region, Andreev bound states (ABSs) are formed, and depend on the superconducting phase difference $\varphi$.
The ABSs are associated with many attractive physics and applications, such as
the $0$-$\pi$ transition~\cite{Ryazanov01},
the $\varphi_0$-state~\cite{Yokoyama14},
the Majorana zero modes~\cite{Lutchyn10},
implementations of superconducting qubit~\cite{Hays21}, etc.

Recently, multi-terminal Josephson junction (MTJJ), where three or more superconducting terminals are connected, has been actively studied.
In MTJJs, multiple macroscopic superconducting phases give rise to nonlocal effects on the Josephson current and the ABSs,
and many studies have been reported e.g., for
nonlocal correlations of supercurrent~\cite{Draelos19}, fractional Shapiro steps~\cite{Matsuo25SS},
anomalous Josephson effect~\cite{Matsuo23AJE,Prosko24}, Josephson diode effect~\cite{Matsuo23JDE,Gupta23},
Andreev molecular states~\cite{Matsuo23AM,Coraiola23AM}, etc.

We study the emergence of Weyl points (WPs)~\cite{Yokoyama15} in the ABS spectrum in MTJJ, which is
a zero-energy degeneracy point with conical dispersion~\cite{Okugawa14}.
The WP acts as a monopole of the Berry curvature being an effective magnetic field.
Then, the WP possesses positive or negative topological charge.
Corresponding to the topological charge, we find the Chern number $N_{\rm Ch}$ as topological invariant by
integrating the Berry curvature on a closed surface.
In Weyl semimetals, the WP leads to unique phenomena such as the negative magnetoresistance~\cite{XHuang15},
the anomalous Hall effect~\cite{Zyuzin12}, and the Nernst effect~\cite{CZeng22}.

In the MTJJs, the WPs emerge in the space spanned by the superconducting phases $\varphi_j$ ($j=1,2,3$ for four-terminal case).
Riwar {\it et al.}, have proposed an experimental scheme to measure the WPs in MTJJs as a transverse conductance owing to the Chern number~\cite{Riwar16,Eriksson17}.
The Chern number $N_{\rm Ch}$ is evaluated by the integral of the Berry curvature in the 2D plane at fixed $\varphi_3$.
By applying finite bias voltages to the first and second superconducting terminals,
$\varphi_1$ and $\varphi_2$ are evolved in time due to the AC Josephson effect and the 2D plane is swept.
This time evolution follows the Chern number $N_{\rm Ch}$ in enough measurement time,
and results in the transverse conductance component in the current through the third terminal, $G_{\rm trans} = -(4e^2/h) N_{\rm Ch}$.
Other detection scheme of the WPs by using microwave spectroscopy has been proposed~\cite{Klees20}.
The MTJJ is a new platform for generating topological effect.
However, the generation and detection of the WPs in experiment are still a challenging issue.

We consider a continuous modulation of the MTJJ.
The normal region of MTJJ should be mesoscopic semiconductor systems, such as quantum well, nanowire, quantum dot, etc.
Such systems can be tuned electrically by the gate electrodes and the number of conduction channels in the terminals affects the physics.
In our previous study, we have modulated the transmission probability of electron channel through the quantum point contacts (QPCs)
and investigated the trajectory of topologically protected WPs due to the system modulation
accompanying with the pair annihilation and creation of WPs.~\cite{Takemura25}.
In this study, we examine an increase of the conduction channels through the QPC structures.
We model the MTJJ by using the random scattering matrix, and consider a statistical average of the WP emergence.
When the number of conduction channels in each terminal is unity, the probability of WP emergence is about 4\%~\cite{Yokoyama15,Riwar16,Takemura25}.
Here, we increase the number of channels up to three, then the emergence probability is enlarged more than 8\% being the double of single channel case.
We examine four-terminal junctions and can assume several cases for the increase of the number of conduction channels in the four terminals.
For the WP emergence, the balance of conduction channels is efficient.

The structure of this paper is as follows. 
In Section 2, we introduce the scattering matrix model with the QPCs and formulate it to include the effects of tuning transmission. 
In Section 3, we discuss the dependence of the WP emergence probability on the number of conduction channels.
Section 4 is devoted to discussion and conclusions.

\section{Model and Formulation}

\begin{figure}[t]
\includegraphics[width=70mm]{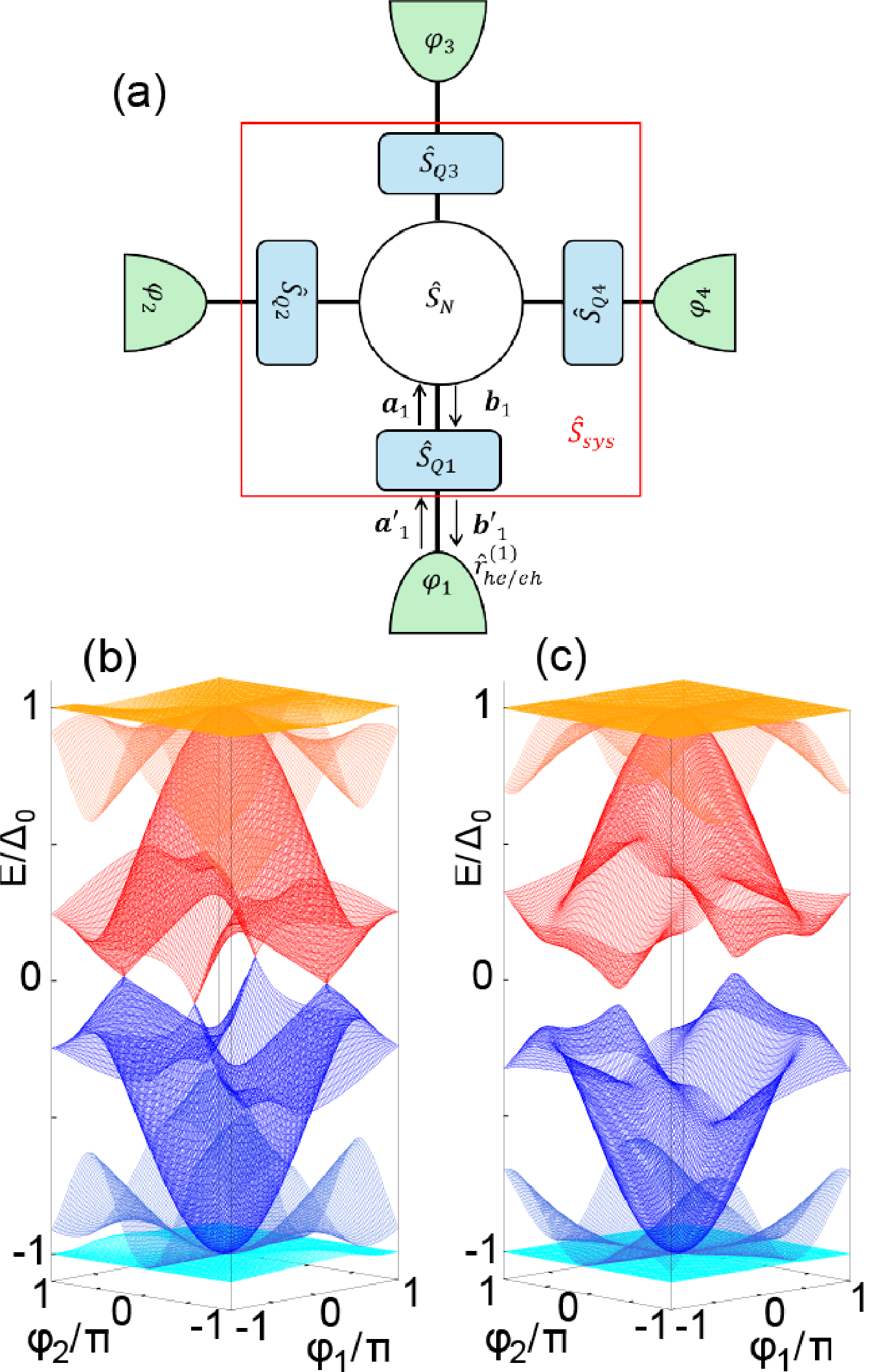}
\caption{Model of the four terminal Josephson junction and the WPs in the energy spectrum of Andreev bound states.
(a) Schematic of the four terminal Josephson junction based on the scattering matrix.
The central block $\hat{s}_{\rm N}$ is the normal region, and the green terminals are superconducting terminals with phase $\varphi_j$.
The QPC structures are embedded between the superconducting terminals and the normal region.
The QPCs are represented by $\hat{s}_{{\rm Q},j}$.
$\hat{s}_{\rm sys}$ is the combined normal region and the QPCs.
$r_{\rm he}^{(1)}$ represents Andreev reflection.
$\bm{a}_j$ ($\bm{a}_j^\prime$) and $\bm{b}_j$ ($\bm{b}_j^\prime$) are the incoming and outgoing waves to the normal region (QPCs), respectively.
(b) A typical result of the Andreev spectrum on a plane crossing the four WPs at $T_j = 0.9$.
$\Delta_0$ is the superconducting gap.
At the positions of WPs,
$(\varphi_1, \varphi_2, \varphi_3) \approx \pm (0.987 \pi,  0.582 \pi, -0.238 \pi)$ and $\pm (0.438 \pi, -0.720 \pi, 0.485 \pi)$,
positive (red) and negative (blue) Andreev levels touch each other at $E=0$.
(c) The Andreev levels at $T_j = 0.2$ when the WPs disappear by the pair annihilation.
Hence, there are only two gapped points.}
\label{fig:model}
\end{figure}

In this section, we introduce the model based on the random matrix and the combination between the QPC structures~\cite{Takemura25}.
First, we explain the scattering matrix model shown in Figure \ref{fig:model}(a).
Then, we describe the method to calculate the ABSs by using Beenakker formula~\cite{Beenakker91}.

\subsection{Scattering matrix description}
Figure \ref{fig:model}(a) schematically shows a model of four-terminal Josephson junction based on the scattering matrix.
The central region described by $\hat{s}_{\rm N}$ corresponds to the normal region.
It is connected to four superconducting terminal with phase $\varphi_j$ ($j=1,2,3,4$).
Such mesoscopic structures can be realized using semiconductor nanocross~\cite{Plissard13}.
We assume the QPC structures between the superconducting terminals and the central normal region.
The QPC potentials are formed by gate voltages on the metallic electrodes.
The electron transmission probability through the QPCs can be tuned electrically and continuously.
The scattering matrices for the QPCs are described by $\hat{s}_{{\rm Q},j}$.
Let us note that the spin degrees of freedom is neglected in the present study.
By the combination of $\hat{s}_{\rm N}$ and $\hat{s}_{{\rm Q},j}$,
we construct the scattering matrix $\hat{s}_{\rm sys}$ of electrons for the mesoscopic system~\cite{Takemura25},
\begin{equation}
\vec{b}_{\rm e}^\prime = \hat{s}_{\rm sys} \vec{a}_{\rm e}^\prime.
\label{eq:Se}
\end{equation}
Here, $\vec{a}_{\rm e}^\prime$ and $\vec{b}_{\rm e}^\prime$ mean the in-coming and out-going wave for electrons, respectively.
The scattering matrix for hole is obtained as $\hat{s}_{\rm sys}^*$,
\begin{equation}
\vec{b}_{\rm h}^\prime = \hat{s}_{\rm sys}^* \vec{a}_{\rm h}^\prime.
\label{eq:Sh}
\end{equation}

At the boundaries between the normal region and the superconducting terminals,
the out-going electron $\vec{b}_{\rm e}^\prime$ (hole $\vec{b}_{\rm h}^\prime$) is reflected to
the in-coming hole $\vec{a}_{\rm h}^\prime$ (electron $\vec{a}_{\rm e}^\prime$).
This process is called the Andreev reflection.
The Andreev reflection is also represented in terms of the scattering matrix as
\begin{equation}
\vec{a}_{\rm h (e)}^\prime = \hat{r}_{\rm he (eh)} \vec{b}_{\rm e (h)}^\prime.
\label{eq:AR}
\end{equation}
The matrix elements of $\hat{r}_{\rm he (eh)}$ are
\begin{equation}
\hat{r}_{\rm he(eh)} = {\rm diag} (r_{1,{\rm he(eh)}}, \cdots , r_{4,{\rm he(eh)}}),
\end{equation}
where $r_{j,{\rm he}} = e^{-i\varphi_j} e^{-i \arccos (E/\Delta_j)}$ and $r_{j,{\rm eh}} = e^{+i\varphi_j} e^{-i \arccos (E/\Delta_0)}$
with $\Delta_j$ being the superconducting gap and $\varphi_j$ being the superconducting phase of the $j$-th terminal.
In the calculation, we set $\varphi_4 = 0$ without loss of generality.

\subsection{Beenakker formulation}

By following the scattering matrices for electron, hole, and the Andreev reflections,
the in-coming wave of electron satisfies
\begin{equation}
\vec{a}_{\rm e}^\prime = \hat{r}_{\rm eh} \hat{s}_{\rm sys}^* \hat{r}_{\rm he} \hat{s}_{\rm sys} \vec{a}_{\rm e}^\prime.
\end{equation}
Then, we obtain the Beenakker formula~\cite{Beenakker91},
\begin{equation}
\det \left( 1 - \hat{r}_{\rm eh} \hat{s}_{\rm sys ,e}^* \hat{r}_{\rm he} \hat{s}_{\rm sys ,e} \right) = 0.
\label{eq:Beenakker}
\end{equation}
Assuming all terminals have the same superconducting gap $\Delta_0$, Eq.\ (\ref{eq:Beenakker}) becomes~\cite{Yokoyama15}
\begin{equation}
\det \left( e^{i 2\chi (E)} - \hat{S} (\vec{\varphi}) \right) =0.
\label{eq:Beenakker2}
\end{equation}
Here, $\chi (E) = \arccos (E/\Delta_0)$ and
\begin{eqnarray}
\hat{S} (\vec{\varphi}) &=& \hat{s}^* (\vec{\varphi}) \hat{s} (\vec{\varphi}),
\label{eq:Smatrix} \\
\hat{s} (\vec{\varphi}) &\equiv & e^{-i \hat{\varphi}/2} \hat{s}_{\rm sys} e^{+i \hat{\varphi}/2}.
\label{eq:smalls}
\end{eqnarray}
$\hat{\varphi}$ means a diagonal matrix by the superconducting phases, $\hat{\varphi} = {\rm diag} (\varphi_1, \varphi_2, \varphi_3 ,\varphi_4)$.

The scattering matrix $\hat{s}_{\rm N}$ for the normal region is given numerically by an orthogonal random matrix~\cite{Yokoyama15}.
By solving Eq.\ (\ref{eq:Beenakker2}), we derive the Andreev levels $E_n (\vec{\varphi})$ of the system.

\subsection{Combination with QPCs}
We construct $\hat{s}_{\rm sys}$ by combining the scattering matrices $\hat{s}_{\rm N}$ for the normal region and $\hat{s}_{{\rm Q},j}$ for the QPCs.
A mathematical detail of the combination is summarized in our previous studies~\cite{Takemura25,Yokoyama17}.
For all incoming and outgoing waves, the equation becomes the following form.
\begin{equation}
\left( \begin{matrix}
\vec{b}_{\rm e}^\prime \\
\vec{a}_{\rm e} \\
\vec{b}_{\rm e}
\end{matrix} \right)
=
\left( \begin{matrix}
\hat{\Sigma}_{11} & 0                           & \hat{\Sigma}_{13} \\
\hat{\Sigma}_{21} & 0                           & \hat{\Sigma}_{23} \\
0                 & \hat{\Sigma}_{32}           & 0
\end{matrix} \right)
\left( \begin{matrix}
\vec{a}_{\rm e}^\prime \\
\vec{a}_{\rm e} \\
\vec{b}_{\rm e}
\end{matrix} \right).
\label{eq:allmatrix}
\end{equation}
The block matrices $\hat{\Sigma}_{\alpha \beta}$ ($\alpha , \beta = 1,2,3$) are given by the elements of $\hat{s}_{\rm N}$ and $\hat{s}_{{\rm Q},j}$.
For example, $\hat{\Sigma}_{32} = \hat{s}_{\rm N}$. After a short algebra, we obtain
\begin{equation}
\hat{s}_{\rm sys} = \hat{\Sigma}_{11} + \hat{\Sigma}_{13} \frac{1}{1 - \hat{\Sigma}_{32} \hat{\Sigma}_{23}} \hat{\Sigma}_{32} \hat{\Sigma}_{21}.
\end{equation}

The size of the scattering matrix $\hat{s}_{\rm N}$ in normal region is $(\sum_{j=1}^4 M_j) \times (\sum_{j=1}^4 M_j)$.
Here, $M_j$ is the number of channels in the $j$-th terminal.
The scattering matrix of the QPC is $2 M_j \times 2 M_j$, which is given as
\begin{equation}
\hat{s}_{{\rm Q},j} 
=\hat{U}_j^\dag 
\left( \begin{matrix}
\hat{t}_{j}^{(1)} &           &0 \\
                            &\ddots & \\
0                          &           & \hat{t}_{j}^{(M_j)}
\end{matrix} \right)
\hat{U}_j .
\label{eq:SQ_Multi}
\end{equation}
Here, $\hat{U}_j$ is $2 M_j \times 2 M_j$ matrix to describe a channel mixing owing to non-adiabatic influence through the QPCs.
$\hat{t}_{j}^{(k)}$ is a $2 \times 2$ matrix to describe an electron transport by the $k$-th channel. 
It is represented as~\cite{NazarovBlanter}
\begin{equation}
\hat{t}_{j}^{(k)}=
\left( \begin{matrix}
\sqrt{1 - t_{j}^{(k)}} e^{i \eta_{j}^{(k)}} & \sqrt{t_{j}^{(k)}}      e^{i \xi_{j}^{(k)}} \\
\sqrt{t_{j}^{(k)}}      e^{i \xi_{j}^{(k)}}    & -\sqrt{1 - t_{j}^{(k)}} e^{i (2\xi_{j}^{(k)} - \eta_{j}^{(k)})}
\end{matrix} \right),
\end{equation}
with the transmission probability $t_j^{(k)}$.
The parameters $\eta_{j}^{(k)}$ and $\xi_{j}^{(k)}$ are phase factors although they are not essential on the Andreev levels.
The transmission of the $j$-th terminal is defined as
\begin{equation}
T_j = \sum_{k=1}^{M_j} t_{j}^{(k)} .
\label{eq:QPC_Trans}
\end{equation}
Note that the increase of conduction channels in the terminals can contribute indirectly to the increase of transverse conductance $G_{\rm trans}$
due to the Chern number $N_{\rm Ch}$ because it relates with the number of WPs.
If eight WPs are emerged, the Chern number can be $N_{\rm Ch} = \pm 2$, then the transverse conductance can also be larger~\cite{Riwar16}.

In a conventional treatment of QPC structure, the electron transport is adiabatic and the channel mixing is not assumed.
However, at out of the conductance plateau, the channel mixing due to impurity and potential roughness might occur~\cite{Zagoskin94,Ulreich98}.
Here, we model and introduce on the scattering matrix.
When the electron transport through the QPCs is adiabatic, the matrix $\hat{U}_j$ in Eq.\ (\ref{eq:SQ_Multi}) is identical matrix.
Let us consider two conduction channels in the $j$-th terminal.
Then, we set $\hat{U}_j$ as
\begin{equation}
\hat{U}_j=
\left( \begin{matrix}
\cos \theta_j   & 0                 & i \sin \theta_j &0                 \\
0                   & \cos \theta_j  & 0                 &i \sin \theta_j \\
 i \sin \theta_j & 0                  & \cos \theta_j  &0                \\
 0                  &i \sin \theta_j & 0                  & \cos \theta_j                
\end{matrix} \right)
\label{eq:Mix_parameter}
\end{equation}
with $\theta_j$ being a channel mixing parameter. At $\theta_j =0$, the transport is adiabatic.
The channel mixing should be weak, hence we consider only small $\theta_j$.

\section{Results and Discussion}

We focus on the emergence of WPs in the Andreev levels.
The scattering matrix $\hat{s}_{\rm N}$ for the normal region is given by the random matrix~\cite{Takemura25}.
In Figs.\ \ref{fig:model}(b) and (c), we demonstrate typical behaviors of the Andreev levels with and without the WPs, respectively,
when the four terminals are $M_j = 1$.
In Fig.\ \ref{fig:model}(b), we select a 2D plane including the WPs and show four conical dispersions at $E=0$ owing to the WPs.
When the transmissions of QPCs are reduced by the gate voltages, the positions of four WPs are shifted and the pair annihilations occur.
Then, the Andreev levels show the gap opening at two positions of the pair annihilations as shown in Fig.\ \ref{fig:model}(c).
In the presence of WPs in Fig.\ \ref{fig:model}(b), the Andreev spectrum possesses finite and discrete Chern number, $N_{\rm Ch}$,
by integrating the Berry curvature on the 2D plane between the WPs~\cite{Riwar16,Eriksson17,Takemura25}.
Then, the presence of WPs can be detected as quantized transverse conductance $G_{\rm trans} = -(4e^2/h) N_{\rm Ch}$.
In Fig.\ \ref{fig:model}(c), the Chern number becomes zero, hence there is no transverse conductance.

In the following, we examine $10,000$ random matrix samples and consider the probability of WP emergence.
We tune the transmission probabilities of electron through the QPCs, $(T_1,T_2,T_3,T_4)$.
The increase of $T_j$ enlarges the WP emergence probability.
We discuss the trend of the WP emergence probability for the observation of WPs in experiment.

Let us note that in the following, $(T_1,T_2,T_3,T_4)=(1,1,1,1)$ is abbreviated as $T(1,1,1,1)$ for readability.

\subsection{Channel Increase}

\begin{figure}[t]
\includegraphics[height=88mm]{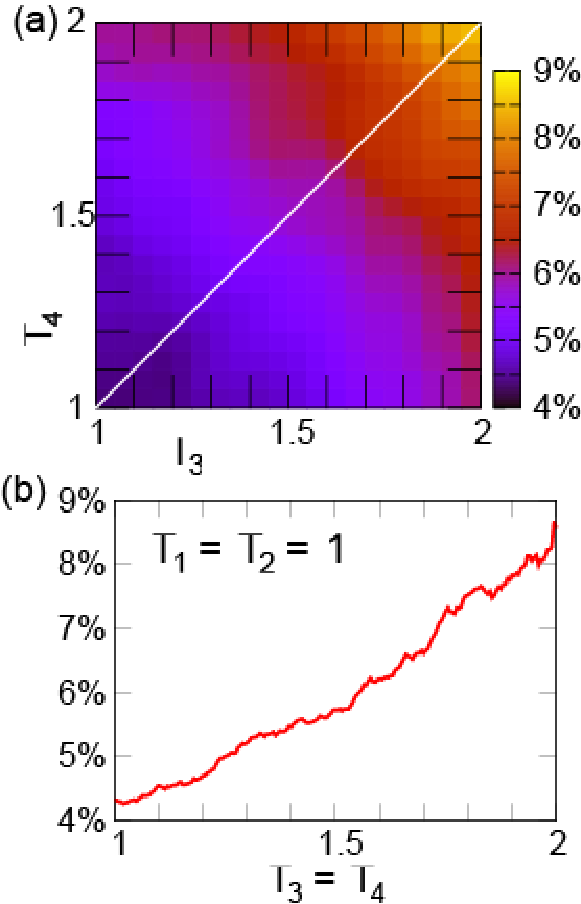}
\caption{Probability of WPs emergence when the QPC transmission probability is changed from $T(1,1,1,1)$ to $T(1,1,2,2)$.
(a) WPs emergence probability map in the $T_3,T_4$ plane at $T_1 = T_2 = 1$.
(b) Cross-sectional plot along the diagonal line in (a).
The WP emergence probability increases from 4.32\% to 8.57\%.}
\label{fig:weyl_prob_map}
\end{figure}

Figure \ref{fig:weyl_prob_map}(a) shows the probability of WP emergence
when the transmissions of the first and second terminals are fixed at $T_1 = T_2 = 1$ and
the third and fourth are varied in the range $1 \le T_3,T_4 \le 2$.
In Fig.\ \ref{fig:weyl_prob_map}(a), the WPs emergence roughly probability is symmetric with respect to the diagonal white line, $T_3 = T_4$.
This symmetry does not hold for individual samples but emerges in the statistics ensembles.
The WP emergence probability increases with the increase of conduction channels in each terminal.
We plot the cross section along the diagonal line in Fig.\ \ref{fig:weyl_prob_map}(b),
hence we tune the system from $T(1,1,1,1)$ (totally four channels) to $T(1,1,2,2)$ (six channels).
The WP emergence probability shows almost linear increase with $T_3 = T_4$.
The probability is about doubly enlarged from 4.32\% to 8.57\%.
Hence, the probability does not linearly increase with the total number of conduction channels.
Note that at $T(1,1,1,1)$, we find only four WPs state, whereas eight WPs state could be found at $T(1,1,2,2)$ although it is very rarely found.
In our discussion, the WP emergence probability considers only existence or absence of the WPs, and the number of WPs is not taken into account.

\begin{figure}[t]
\includegraphics[height=88mm]{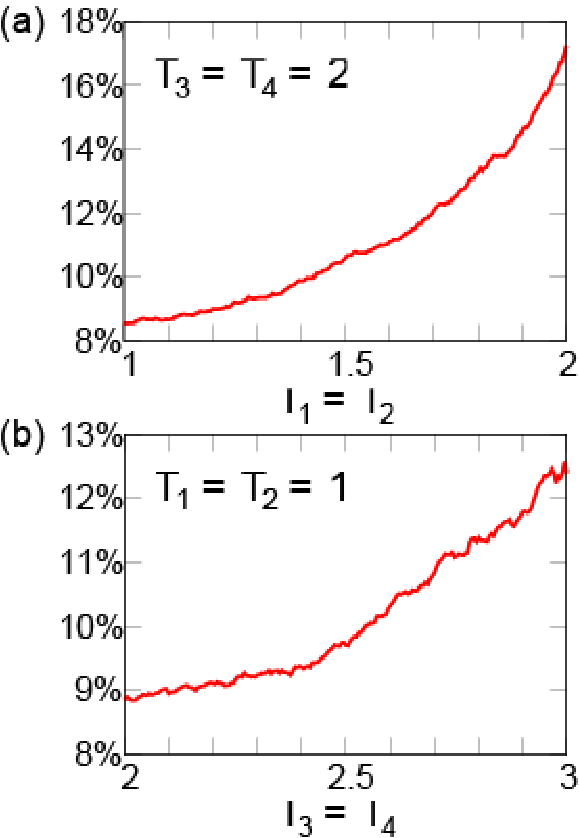}
\caption{Probability of WPs emergence when the total number of conduction channels in systems increases from six to eight.
(a) From $T(1,1,2,2)$ to $T(2,2,2,2)$ with $T_1 = T_2$.
The WP emergence probability varies from 8.57\% to 17.24\%.
(b) From $T(1,1,2,2)$ to $T(1,1,3,3)$ with $T_3 = T_4$.
The WP emergence probability varies from 8.85\% to 12.4\%.}
\label{fig:weyl_prob_8channel}
\end{figure}

In Fig.\ \ref{fig:weyl_prob_8channel}, we examine the WP emergence probability from total six conduction channels to eight channels.
Figure \ref{fig:weyl_prob_8channel}(a) shows the increase of WP emergence probability from $8.57\%$ at $T(1,1,2,2)$ to $17.24\%$ at $T(2,2,2,2)$.
Note that at $T(1,1,2,2)$ in (a), (b), and in Fig.\ \ref{fig:weyl_prob_map}, the statistical probabilities are not identical.
It is owing to the different size of scattering matrices, and not essential in the present discussions.
In this parameter range, the WP emergence probability is deviated from the linear increase.
The probability at $T(2,2,2,2)$ is enlarged up to about four times than that of $T(1,1,1,1)$
although the total number of conduction channels increase only doubly from four to eight.
Hence, the WP emergence probability per conduction channels also increase.
This suggests that the increase of total conduction channels might be advantage to emerge the WPs.
In addition, the number of emergent WPs is also increased.
However, enormous number of the conduction channels results in enormous number of the Andreev bound states,
which disturbs to distinguish the conical dispersion of the WPs owing to the continuous density of states at zero energy~\cite{Yokoyama17}.

We examine the other distribution of conduction channels.
In Fig.\ \ref{fig:weyl_prob_8channel}(b), the conduction channels are tuned from $T(1,1,2,2)$ to $T(1,1,3,3)$.
Then, the WP emergence probability increases up to only $12.4\%$, which about three time of that at $T(1,1,1,1)$
and is significantly smaller than that at $T(2,2,2,2)$.
Hence, a balanced distribution of conduction channels to the superconducting terminals is beneficial.

\begin{figure}[t]
\includegraphics[width=85mm]{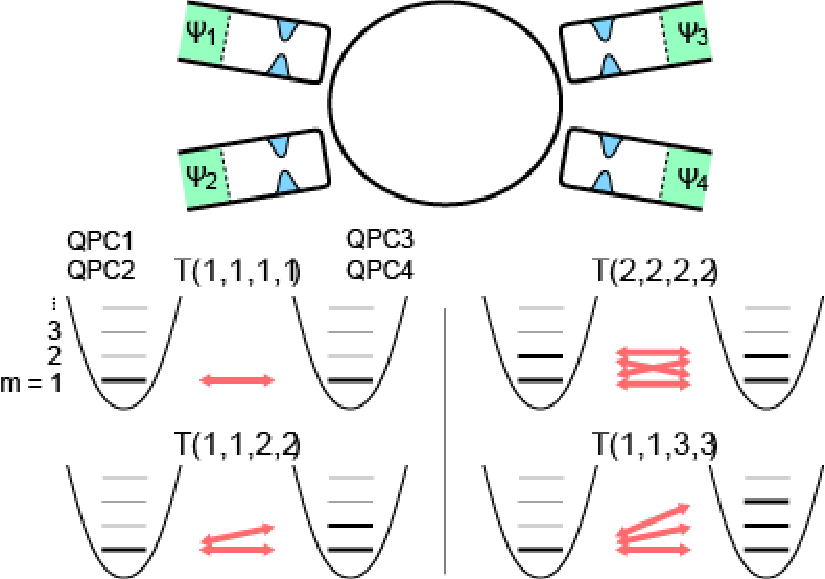}
\caption{Schematic illustration for the combinations of conduction channels.
As a simplified diagram for ease of understanding, the QPC 1, 2 and the QPC 3, 4 are unified, respectively.}
\label{fig:combination}
\end{figure}

In our results, $T(1,1,1,1)$ has 1 combination and an emergence probability of $4.32\%$; $T(1,1,2,2)$ has 4 combinations
and $8.57\%$; $T(2,2,2,2)$ has 16 combinations and $17.24\%$; and $T(1,1,3,3)$ has 9 combinations and $12.4\%$.
The emergence of WPs requires four superconducting terminals~\cite{Yokoyama15}, hence $T(1,1,1,1)$ is minimal situation for WP.
At $T(1,1,2,2)$, in addition to the lowest channels ($m=1$) in the third and fourth terminals,
the second lowest channels ($m=2$) could cooperate with the lowest channels ($m=1$) in the first and second terminals,
which increases doubly the emergence probability.
We illustrate schematic explanations in Fig.\ \ref{fig:combination}.
In the case of $T(1,1,3,3)$, the third lowest channels ($m=3$) in the third and fourth terminals contributes to the WP emergence with
the lowest channels ($m=1$) in the first and second terminals.
In this case, the third channels ($m=2,3$) in the third and fourth terminals do not cooperate with $m=1$ and $2$ in the same terminals.
For the case of $T(2,2,2,2)$, $m=2$ channels can cooperate with themselves and $m=1$ channels.
Then, we obtain four times enhancement of the WP emergence probability from $T(1,1,1,1)$ to $T(2,2,2,2)$.
Therefore, the balance of QPCs to maximize the combination of channels is most effective for WP emergence rather than the total number of channels.

\subsection{Channel Mixing}

The increase of conduction channels results in the enhancement of WP emergence probability
owing to the increase of the number of channel combinations for the Andreev bound states.
Then, we examine the influence of conduction channel mixing, which might contribute to the channel combination.
Generally, such effect due to the channel mixing is considered in the scattering matrix $\hat{s}_{\rm N}$.
Moreover, the electron transport through the QPC structures could be adiabatic in theory.
In experiment, however, non-adiabatic process might occur at the QPCs.
Here, we assume a slight channel mixing at the QPCs and examine the tuning of channel mixing in the mesoscopic systems, $\hat{s}_{\rm sys}$.

\begin{figure}[t]
\includegraphics[width=50mm]{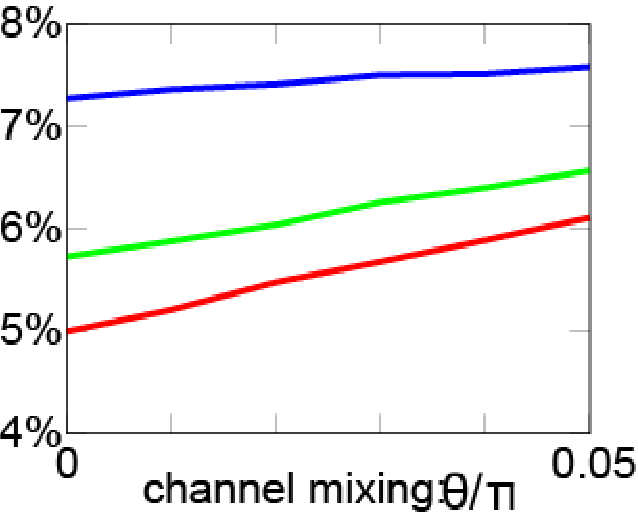}
\caption{WP emergence probability when a weak channel mixing due to non-adiabatic transport through the QPCs is considered at
$T(1,1,T_3,T_4)$ with $T_3=T_4=1.25$ (red), $1.50$ (green), and $1.75$ (blue lines).
The channel mixing parameter is $\theta = \theta_3 = \theta_4$.}
\label{fig:Mix_theta}
\end{figure}

Figure \ref{fig:Mix_theta} shows an increase of the WP emergence probability with the increase of channel mixing parameter $\theta$.
At the conductance plateau in the QPCs, the channel mixing might be insufficient.
Here, we consider the mixing in the third and fourth terminals at out of the conductance plateau,
$T_3=T_4=1.25$ (red), $1.50$ (green), and $1.75$ (blue lines). The other QPCs are set at $T_1 = T_2 = 1$.
A correlation between the mixing strength $\theta$ and the WP emergence probability is positive and
is stronger at lower electron transmissions through the QPCs.
It can be explained by the combination of channels involved in WP emergence, as in the previous subsection.

\section{Conclusions}
\label{sec:conclusions}

We have investigated the effect of the number of conduction channels on the emergence of WPs in four terminal Josephson junctions based on the scattering matrix.
The number of channels and their transmission probabilities are tuned by the QPC structures at the terminals.
We have examined the statistic analysis for many samples of random matrices and discussed the emergence probability of WPs.
The increase of the total number of conduction channels in the systems results in the increase of the WP emergence probability.
Then, the balanced distribution of channels is efficient for the WP emergence.
Especially, the emergence probability for the case of $T(2,2,2,2)$ is $4/3$ times larger than that of $T(1,1,3,3)$ even though the total channel number is the same.
This result suggests that the number of combinations of channels in the four terminals contributes linearly the WP emergence.
To support it, we examine an artificial weak modulation of the conduction channel mixing due to the non-adiabatic electron transport through the QPCs,
and we find that the channel mixing increases the WP emergence probability.

In the present discussion, we restrict the number of conduction channels in each terminal is less than four.
In our framework, we can increase it up to more than 100 easily.
However, in such enormous channels, the Andreev spectrum becomes quasi-continuous and the density of states at zero energy becomes finite~\cite{Yokoyama17}.
Then, it is difficult to distinguish the WP dispersions at zero energy.
Hence, we consider only a few conduction channels in the terminals.

We can extend the present framework to include the effect of channel mixing due to the spin-orbit interaction~\cite{Eto05}.
It may offer additional handles for controlling WP emergence as our future work.

\begin{acknowledgments}
K.T. is supported by JST, the establishment of university fellowships towards the creation of science technology innovation,
Grant Number JPMJFS2125.
We thank Prof. Akira Oiwa at Osaka University for a fruitful discussion.
\end{acknowledgments}

\end{document}